\documentclass[11pt,dvips]{article}
\usepackage{graphicx}
\begin{document}

\large \baselineskip4ex

\begin{center}
{\noindent \Large\bf Electronic Structure and Magnetism of Equiatomic FeN \\} \vspace{0.2in}
 {\Large Y. Kong\\ \vspace{0.3cm}{\normalsize {\em Department
of Physics {\sl \&} The Applied Magnetics Laboratory of the Ministry \\ of
Education, Lanzhou University, 730000 Lanzhou, China\\ Max-Planck-Institut
f\"{u}r Festk\"{o}rperforschung, Heisenbergstr. 1, D-70569 Stuttgart,
Germany\\}}  }

\vspace{0.3in}{\Large\bf Abstract}
\end{center}

\large \baselineskip4ex

In order to investigate the phase stability of equiatomic FeN compounds
and the structure-dependent magnetic properties, the electronic structure
and total energy of FeN with NaCl, ZnS and CsCl structures and various
magnetic configurations are calculated using the first-principle
TB-LMTO-ASA method. Among all the FeN phases considered, the
antiferromagnetic (AFM) NaCl structure with $q=(0,0,\pi)$ is found to have
the lowest energy at the theoretical equilibrium volume. However, the
ferromagnetic (FM) NaCl phase lies only 1mRy higher. The estimated
equilibrium lattice constant $a^{th}$=4.36{\AA} for nonmagnetic (NM)
ZnS-type FeN agrees quite well with the experimental value of
$a^{exp}$=4.33{\AA} but for AFM NaCl phase the $a^{th}$=4.20{\AA} is 6.7\%
smaller than the value observed experimentally. For ZnS-type FeN,
metastable magnetic states are found for volumes larger than the
equilibrium value. With the analysis of atom- and orbital-projected
density of states (DOS) and orbital-resolved crystal orbital Hamilton
population (COHP) the iron-nitrogen interactions in NM-ZnS, AFM-NaCl and
FM-CsCl structures are discussed. The leading Fe-N interaction is due to
the $d$-$p$ iron-nitrogen hybridization while considerable $s$-$p$ and
$p$-$p$ hybridizations are also observed in all three phases. The iron
magnetic moment $\mu_{Fe}$ in FeN is found to be highly sensitive to the
nearest-neighboring Fe-N distance. In particular, the $\mu_{Fe}$ in ZnS
and CsCl structures show an abrupt drop from the value of about 2$\mu_{B}$
to zero with the reduction of the Fe-N distance.

~

\newpage
\large \baselineskip4ex

\section{Introduction}
\hspace{20pt}For a long time iron nitride has attracted much scientific
interest in basic research as well as in technology-oriented research.
While a large number of results for iron-rich nitrides, such as
$\gamma^{\prime}$-Fe$_4$N, have appeared in the literature\cite{mohn},
only a few investigations on high N-content FeN, specifically
FeN\cite{heiman, bauer} with the proportionality 1:1 of Fe and N atoms
were reported. Recently, two kinds of FeN structures have been determined
in FeN films\cite{nakagawa,suzuki,rissanen} synthesized by sputtering
technique. They are the sodium chloride structure with lattice constants
a=4.5{\AA} and the zinc blende structure with lattice constant
a$\approx$4.33\AA. All the prepared FeN films were non-stoichiometric due
to N vacancies and/or impurities. $^{57}$Fe M\"{o}ssbauer spectroscopy
experiments\cite{nakagawa, hinomura} have shown that at 4.2K no magnetic
hyperfine splitting is observed for ZnS-type FeN, but two kinds of Fe
sites in NaCl-type FeN exhibit surprisingly large hyperfine fields of 49T
and 30T. It was suggested that the ZnS-type FeN is nonmagnetic (NM) while
the NaCl-type FeN shows antiferromagnetic (AFM)
coupling\cite{nakagawa,hinomura}. However, the magnetic measurements by
Suzuki {\sl et al.}\cite{suzuki} indicated that at low temperature the
stoichiometric ZnS-type FeN is AFM and exhibits a micromagnetic character.

\par Using a full-potential linearized-augmented-plane-wave (FLAPW) method,
Shimizu {\sl et al}.\cite{shimizu1,shimizu} calculated the electronic,
structural and magnetic properties of stoichiometric ZnS- and NaCl-type
FeN and estimated the equilibrium lattice constants and bulk moduli. The
results indicated that the ferromagnetic (FM) NaCl-type FeN was more
stable than the NM-ZnS structure. Furthermore, the calculated
results\cite{shimizu} for the NaCl-type FeN with NM, FM, $(\pi,\pi,\pi)$-
and $(0,0,\pi)$-AFM configurations identified that the FM structure was
the ground state with equilibrium lattice constant $a^{th}\sim$4.0\AA. It
has further shown that at the experimental lattice constant 4.5{\AA} the
$(\pi,\pi,\pi)$-AFM state was more stable than the FM state. A similar
detailed analysis was not carried out for the ZnS-type FeN and the reason
for the difference in magnetism between the two structures was therefore
not clear. Most recently, Eck {\sl et al}.\cite{eck} performed band
structure calculations using tight-binding-linear-muffin-tin-orbitals
(TB-LMTO) method within the atomic-sphere-approximation (ASA) and
investigated structural and electronic properties of both FeN structures
by analyzing density of states (DOS) and crystal orbital Hamilton
population (COHP). They concluded that the zinc blende structure should be
more stable because of the weaker antibonding Fe-Fe interactions below the
Fermi level. Moreover, the authors evaluated the theoretical lattice
constants for non-stoichiometric FeN and found that the deficiency of N
atom decreases the equilibrium lattice constant. The magnetic properties
of FeN were however hardly discussed. To achieve a better understanding of
the magnetic diversity in FeN, systematic investigation of the electronic
structure and magnetic properties of various FeN structures thus seems to
be needed.

\par In this paper, the TB-LMTO-ASA method is applied to calculate band
structures of FeN compounds with various structures and spin
configurations. The structural, electronic and magnetic properties of the
FeN are thus investigated by analyzing the calculated electronic structure
and total energy. In addition to the NaCl and ZnS structures, we also
consider FeN in the CsCl structure.

\par A brief description of computational technique will be given in the
next section. In section 3, we present the calculated total energy and
electronic structural and discuss the structure and magnetic properties of
FeN compounds. A short summary is given in the last section.

\section{Computational Details}
\hspace{20pt}In the present study the electronic structures of equiatomic
FeN compounds are calculated self-consistently using the
scalar-relativistic TB-LMTO method\cite{lmto} in the
atomic-sphere-approximation including the combined correction. A
local-density-approximation (LDA)\cite{exco} to the density-functional
theory (DFT) is applied  and the non-local correction is also included
using the Perdew-Wang generalized-gradient-approximation (GGA).\cite{GGA}

\par The lattice structures of NaCl-, ZnS- and CsCl-type FeN compounds are
illustrated in Fig. 1. The Fe atoms in sodium chloride and zinc blende
structures form an fcc lattice with different coordination of the N atoms.
In the NaCl structure the N atoms are located at octahedral sites of the
fcc lattice and each Fe atom has six N neighbors while in the ZnS
structure the half-filling of the N atoms in tetrahedral sites of the fcc
lattice makes each Fe four-fold coordinated to N atoms. In the cesium
chloride structure the N atoms fill the body-centered sites of a
simple-cubic Fe lattice and one Fe is eight-fold coordinated to N. The
sites and filling fractions of the N atoms, the experimental magnetic
structure, the nearest-neighbor (NN) Fe-Fe and Fe-N distances as well as
the number of N neighbors of an Fe atom are summarized for the different
structures in Table 1.

\par We perform NM, FM and ($0,0,\pi$)-AFM calculations for FeN with NaCl, ZnS
and CsCl structures. Besides $(0,0,\pi)$-AFM state, the
$(\pi,\pi,\pi)$-AFM configuration and an additional AFM
structure\cite{entel}, which consists of double layers with FM interlayer
coupling along [001]-direction, AFMD, are also considered for NaCl-type
FeN to include possible magnetic coupling in FeN film. The $k$-space
integration is performed with the tetrahedron method\cite{integration}
using $16\times 16\times 16$ mesh within the Brillouin zone. All partial
waves with $l\leq 2$ are included in the basis for Fe as well as N. For
ZnS- and NaCl-type FeN, special care is taken in filling the interatomic
space since the structures are rather open. Therefore, it is necessary to
introduce interstitial spheres. The sphere radii and the positions of
interstitial spheres are chosen in such a way that space filling is
achieved without exceeding a sphere overlap of 16\%. This is done using an
automatic procedure developed by Krier {\sl et al}.\cite{Krier}. The radii
of atomic and interstitial spheres corresponding to the theoretical
equilibrium volume are listed in Table 2 for the different FeN structures.

\par In combination with the analysis of atom- and orbital-projected
density of states (DOS), the COHP technique\cite{cohp} is applied to analyze
the chemical bonding in ZnS-, NaCl- and CsCl-type FeN to examine the
different Fe-N interactions. This technique provides information analogous
to the Crystal Orbital Overlap Population (COOP) analysis.\cite{coop}
While COOP curves are energy resolved plots of the Mulliken overlap
population between two atoms or orbitals, a COHP curve is an energy
resolved plot of the contribution of a given bond to the bonding energy of
the system. Similar to COOP curves, in all the COHP curves presented here
positive values are bonding and negative antibonding, i.e. $-$COHP is
plotted instead of COHP.

\section{Results and Discussions}
\subsection{Total energy and phase stability}
\hspace{20pt}The calculated total-energy curves for ZnS-, NaCl- and
CsCl-type FeN with NM, FM and AFM spin configurations are shown in Fig.
2. The theoretical equilibrium lattice constants $a^{th}$ are estimated
for various FeN phases and they are listed in Table 2 together with total
energies at $a^{th}$. The total-energy for NM CsCl structure is not
plotted in Fig. 2 because the values are much higher than the other
phases.

\par Firstly, we shall discuss the total-energy curves for NaCl-type FeN.
Except for the NM NaCl-type FeN with higher energy, the magnetic NaCl
phases with different spin configurations give similar total-energy
curves. As indicated in Table 2, at the equilibrium volume the $E^{th}$
for ($0,0,\pi$)-AFM phase is only 1mRy lower than that for FM-NaCl
structure and 10mRy below the highest ($\pi,\pi,\pi$)-AFM structure.
Although our results are in better agreement with experiment than previous
results\cite{shimizu}, the $a^{th}$ ($\sim$4.20\AA) estimated for
NaCl-type FeN is still 6.7\% smaller than the values observed
experimentally. This discrepancy is much larger than the normal deviation
due to the local density approximation (LDA) with GGA correction. It may
therefore be, as suggested in Refs.\cite{shimizu,eck},  that the
experimental NaCl-type FeN with $a$=4.5{\AA} is not in stable state due to
unknown effects, such as the effect of the surface.

\par We also perform total-energy calculations
for magnetic NaCl phases without the inclusion of the GGA correction to
the LDA. The estimated theoretical lattice constants for magnetic NaCl
phases are about 3.96$\sim$4.01{\AA}, which are smaller than those
obtained with GGA, and the FM phase is found to be stable at the
theoretical equilibrium volume. The results are consistent with those by
Shimizu {\sl et al}.\cite{shimizu}. In the following only the results
calculated with the GGA correction are presented.

\par For ZnS-type FeN, a NM ground state is found with theoretical
equilibrium lattice constant $a^{th}$=4.36{\AA}, which agrees quite well
with the experimental value\cite{nakagawa}. The state is about 6mRy higher
in energy than that of the ($0,0,\pi$)-AFM NaCl structure. Self-consistent
($0,0,\pi$)-AFM and FM solutions are found when the lattice constant is
larger than 4.45{\AA} and 4.69{\AA}, respectively. Since the critical
lattice constant 4.45{\AA} for the ($0,0,\pi$)-AFM state is only 0.09{\AA}
larger than $a^{th}$ of the NM-ZnS phase, one could expect that the
metastable AFM phase would be important for the observed micromagnetic
character in ZnS-type FeN.\cite{suzuki}

\par The total-energy curves calculated for FM and ($0,0,\pi$)-AFM
CsCl-type FeN are nearly identical and about 70mRy above those of ZnS and
NaCl structures. Because of this rather higher total-energy, the
CsCl-structure should not be considered a possible stable FeN phase.

\par Among all the investigated FeN phases, as mentioned above, the simple
($0,0,\pi$)-AFM NaCl structure is found to have the lowest total energy at
theoretical equilibrium volume, while the $E^{th}$ for FM-NaCl structure
is only about 1mRy higher. However, the energy difference is very small
and may be of the order of the inaccuracies due to the ASA used in our
calculations.

\subsection{Electronic structure and iron-nitrogen interactions}
\hspace{20pt}In the following, the electronic structure and Fe-N
interactions in the FeN phases will be analyzed in detail using the
calculated DOS and COHP for NM ZnS, ($0,0,\pi$)-AFM NaCl and FM CsCl
structures. In Fig. 3 we show the spin-resolved DOS for NM ZnS-,
($0,0,\pi$)-AFM NaCl- and FM CsCl-type FeN at the theoretical equilibrium
volume. The DOS at Fermi level, $N(E_F)$, are listed in Table 2.

\par Except for the $s$-like states around $-15$eV, which are not shown,
the DOS of NM-ZnS structure shown in Fig. 3a mainly consists of two sets
of structures, which are separated by a 2eV energy gap. The lower energy
set, centered around $-6$eV, is primarily composed of N 2$p$ bands with
some admixture of Fe $d-$character, the higher energy set from $-2.3$ to
$3.6$eV, is dominated by Fe 3$d$ states. A small gap at $-0.5$eV further
divides the latter into two parts, which are respectively characterized by
the crystal-field splitted $e$ and $t_2$ sets. The Fermi-level $E_F$ is
located just above the gap at the low-energy side of the $t_2$ set. The
reason for ZnS-type FeN phase showing no magnetic moment at equilibrium
volume may be understood from the DOS at the $E_F$. According to the
Stoner model the self-consistency condition for a ferromagnet can be
simply expressed by $$N(E_F){\cdot}I=1$$ where $N(E_F)$ is the
paramagnetic density of state at the $E_F$ and $I$ the effective Stoner
interaction parameter. For the ZnS-type FeN, it is observed that the DOS
curve intersects the Fermi-level with a large negative slope and gives a
quite small $N(E_F)$ (0.7 states/eV$\cdot$spin). Consequently, no magnetic
ordering is observed for ZnS-type FeN. On the other hand, by increasing
the volume, the 3$d$ subband becomes much narrower and the $N(E_F)$
larger. The Stoner criterion may therefore be satisfied and the ZnS-type
FeN shows certain magnetic ordering. Indeed, our total-energy calculations
for ZnS-type FeN has confirmed the metastable FM and AFM phases existing
at larger volumes.

\par Since the band structures of two sublattice in AFM phase are
identical, in Fig. 3b only the sublattice DOS for ($0,0,\pi$)-AFM NaCl
structure is plotted. Compared to that of NM ZnS phase, the DOS of
($0,0,\pi$)-AFM NaCl-type FeN form a continuous spectrum over the energy
range between $-8$ and $4$eV. Due to the large dispersion of Fe 3$d$
subbands the expected crystal field splitting of 3$d$ orbitals can not be
clearly seen. While the majority-spin DOS are nearly completely occupied,
a dominant minority-spin peak is located above the Fermi level due to the
exchange-splitting. In Fig. 3c the peaks of the DOS for FM CsCl structure
also show complicated structure. Compared to that of NaCl structure, the
DOS of the CsCl-type FeN exhibits larger dispersion. Below the $E_F$, the
majority-spin DOS peaks of NaCl and CsCl structures have a tendency to
split into two parts by a deep valley at about $-3$ and $-4$eV,
respectively. The higher part is dominated by Fe 3$d$ states while the
lower one is mostly composed of N 2$p$ with some admixture of Fe $d$
states.

\par The partial DOS (PDOS) of FeN with NM ZnS, ($0,0,\pi$)-AFM NaCl and FM
CsCl structures, projected for Fe $s$, $p$, $d$ and N $p$ orbitals, are
plotted in Fig. 4. Correspondingly, Fig. 5 shows the orbital-projected
COHP calculated for the $s$-$p$, $p$-$p$ and $d$-$p$ iron-nitrogen
interactions. Here only the nearest-neighbor contributions are included in
the COHP. Noticed that of the two identical sublattices in
($0,0,\pi$)-AFM-NaCl phase, only the PDOS for one sublattice are shown in
Fig. 4b while the COHP curves for Fe$_1$-N$_1$ (Fe$_2$-N$_2$)
intra-sublattice and Fe$_1$-N$_2$ (Fe$_2$-N$_1$) inter-sublattice
interactions are presented in Fig. 5b and 5c, respectively. Here Fe$_1$
(N$_1$) defines the Fe (N) atom in the up-polarized sublattice and Fe$_2$
(N$_2$) the atom in the down-polarized sublattice.

\par Consistent with the observation of the total DOS, a 2eV energy gap
separates the PDOS of NM ZnS phase, shown in Fig. 4a, into two sets of
isolated structures. As indicated by the COHP curves in Fig. 5a, N $p$
orbitals mix Fe $s, p$ and $d$ subbands on both side of the gap but the
$d$-$p$ hybridization dominates the iron-nitrogen interactions. The
$d$-$p$ Fe-N hybridization is characterized by the large bonding peaks
below the gap, centered at about $-5$eV, and the antibonding peaks around
the Fermi-level, centered at about 2.0eV. On the contrary, in both energy
windows the $s$-$p$ and $p$-$p$ hybridizations form bonding interactions.
The corresponding antibonding interactions lie far above the Fermi-level.

\par In ($0,0,\pi$)-AFM NaCl structure, the PDOS projected for Fe $s, p, d$
and N $p$ orbitals (Fig. 4b) spread over the same energy range from $-8$
to 4.0eV without energy gap and show prominent hybridizations in the
entire energy range. According to the calculated COHP shown in Fig. 5b and
Fig. 5c, the $d$-$p$ hybridization in NaCl-type FeN is characteristic of
the bonding interactions below $-3$eV and the antibonding states above
$-3$eV. Furthermore, it is found that the antibonding peaks below the
$E_F$ are dominated by antibonding $\pi^*$ states and the peaks above the
$E_F$ are mainly $\sigma^*$ states in character. Comparing the $d$-$p$
interactions in Fig. 5b and Fig. 5c, a stronger antibonding $\pi^*$ peak
at $-2$eV is observed for inter-sublattice Fe-N hybridization. Except for
the leading $d$-$p$ interactions the $s$-$p$ and $p$-$p$ hybridization
show considerable bonding interactions from $-8$ to 4eV. In contrast to
the $d$-$p$ interaction, the $s$-$p$ and $p$-$p$ interactions for both
intra- and inter-sublattice Fe-N interactions are nearly identical.
Compared to the ZnS structure, the antibonding peaks of $s$-$p$ and
$p$-$p$ hybridization in NaCl structure are shifted towards lower energy.

\par Due to larger Fe-N distance, the Fe-N interactions in FM CsCl
structure is much weaker than that in ZnS and NaCl structure. As also
observed in ($0,0,\pi$)-AFM NaCl structure, the dominant Fe-N $d$-$p$
hybridization is composed of the bonding set at lower energy and the
antibonding set around the $E_F$. Below the $E_F$ there exist strong
antibonding peaks. For $s$-$p$ and $p$-$p$ interactions the antibonding
peaks are found to be much lower in energy than those in ZnS and NaCl
structures.

\par Although the $s$-$p$ and $p$-$p$ hybridizations give inevitable
contributions, as indicated by the COHP curves in Fig. 5, the dominant
Fe-N interaction in all the three FeN structures is the $d$-$p$
hybridization. In Fig. 5 we also plot the integrated COHP for
$d$-$p$ interactions. As indicated by the ICOHP at Fermi level, the
ZnS-type FeN exhibits the strongest $d$-$p$ Fe-N hybridization among the
three FeN structures because of the shortest nearest-neighboring Fe-N
distance. On the contrary, the largest Fe-N distance in CsCl structure
makes the CsCl-type FeN showing the weakest Fe-N hybridization although
the Fe atoms in this structure has the most N neighbors.

\subsection{Magnetic moment and its volume dependence}
\hspace{20pt}The calculated iron magnetic moments $\mu_{Fe}$ for the
different FeN phases at the theoretical equilibrium volume are listed in
table 2. For all the four magnetic NaCl-type FeN phases a similar value of
$\mu_{Fe}$ is obtained. Among them the FM phase produces a slightly larger
$\mu_{Fe}$. For both FM and AFM CsCl-type FeN the calculated Fe magnetic
moments are larger than those in NaCl-type FeN. At equilibrium volume no
spin-splitting for FeN with ZnS structure is obtained from spin-polarized
FM and AFM calculations. Noticed that the Fe-N interaction in ZnS
structure is the strongest and that in CsCl structure the weakest, it is
reasonably concluded that the magnetic properties of Fe in FeN structures
is strongly correlated to the strength of the Fe-N interactions, mainly
$d$-$p$ hybridization. In other words, the magnetic properties of FeN
phases is closely related to the nearest-neighboring Fe-N distance in the
structure. The stronger the Fe-N hybridization is, the smaller becomes the
Fe magnetic moment.

\par The Fe magnetic moments in magnetic NaCl-, CsCl- and ZnS-type FeN
calculated at various volumes are shown in Fig. 6 as a function of the
nearest-neighboring Fe-N distance, $d_{Fe-N}$. It is found that the
$\mu_{Fe}$ is highly sensitive to the $d_{Fe-N}$. By compressing the
volume, the Fe magnetic moment in all the magnetic FeN phases dramatically
decrease with the reduction of the $d_{Fe-N}$. While $\mu_{Fe}$ in
NaCl-type FeN decrease with $d_{Fe-N}$ at nearly the same rate and it
becomes zero gradually when the $d_{Fe-N}$ is smaller than about 3.3{\AA},
$\mu_{Fe}$ in both CsCl and ZnS structures exhibit a sudden drop from a
value of about 2$\mu_B$ to zero at a certain critical $d_{Fe-N}$.
According to the canonical band model\cite{andersen}, the different volume
dependence of the $\mu_{Fe}$ in various FeN structures should be related
to the difference of the DOS around the $E_F$.

\section{Conclusions}
\hspace{20pt}Starting out from the calculated total energy and electronic
structure using the TB-LMTO-ASA method, we have investigated structural,
electronic and magnetic properties of FeN compounds with NaCl, ZnS and
CsCl structures.

\par From the calculated total-energy a stable ($0,0,\pi$)-AFM NaCl-type FeN
with theoretical equilibrium lattice constant $a=4.2${\AA} is identified.
For the FeN with ZnS structure, our results indicate the existence of
metastable AFM and FM solutions with a larger cell volume besides stable
NM phase. These metastable states may be important for the observed
micromagnetic character in ZnS-type FeN.

\par By analyzing the atom- and orbital-resolved density of states and the
orbital-projected COHP, the Fe-N hybridization interactions in NM ZnS,
($0,0,\pi$)-AFM NaCl and FM CsCl structures are discussed in detail. In
all the three considered FeN structures the $d$-$p$ hybridization between
Fe and N atoms dominates the Fe-N interactions. With the calculated Fe
magnetic moment for various FeN structures, it is suggested that the
magnetic properties of FeN phases are closely related to the strength of
the Fe-N $d$-$p$ hybridization. Furthermore, the Fe magnetic moment in FeN
phases are found to be highly sensitive to the nearest-neighboring Fe-N
distance.

\par Finally, it should be mentioned that the equilibrium lattice
constant $a^{th}$ estimated for NaCl-type FeN is 6.7\% smaller than the
values observed experimentally though in our calculations the GGA
correction to the local density approximation has been applied. It is
suggested that other unknown effects should be responsible for the
extraordinarily larger lattice constant observed for FeN with NaCl
structure. Eck et al.\cite{eck} have estimated the equilibrium lattice
constant for defected FeN and obtained a rather smaller $a^{th}$. Up to
date, no theoretical investigations on FeN film has been reported. It is
highly desired to perform calculations on FeN structure with
low-dimensional symmetry so that the effect of surface may be examined.

\section*{Acknowledgements}
The author would like to gratefully acknowledge Prof. O.K. Andersen for
many helpful advices and is indebted to Dr. O. Jepsen for a careful
reading of the manuscript.

\newpage
\section*{Tables}
\begin{table}[h]\caption{\label{structure}The site and filling fraction (per
cubic cell) of the N atoms in FeN compounds, lattice constant $a$ and
magnetic structures. $d_{Fe-Fe}$ and $d_{Fe-N}$ are the nearest-neighbor
Fe-Fe and Fe-N distances at the experimental lattice constant. $n$ is the
number of N neighbors of one Fe atom. Since the CsCl-type FeN does not
exist, the calculated results are given. }
\begin{center}
  \begin{tabular}{ccccc}\hline \hline
Structure type& &ZnS &NaCl &CsCl \\ \hline
Site& & tetrahedral& octahedral &body-centered \\
filling fraction & &4/8&4/4&1/1 \\
exp. mag. structure & &NM&AFM &FM  \\
$a$ (\AA) & & 4.33\cite{nakagawa} & 4.50\cite{nakagawa} &2.63 \\
$d_{Fe-Fe}$ (\AA)& &3.04&3.18&2.63 \\
$d_{Fe-N}$ (\AA) & &1.86&2.25&2.28 \\
$n$ &     & 4   &  6    & 8  \\ \hline
  \end{tabular} \\
{}\end{center}
\end{table}

\begin{table}[h]\caption{Calculated total energy, equilibrium lattice
constant, DOS at $E_F$ and Fe magnetic moment for some considered FeN
structures. E$^{th}$ is the total energy (given by E+2656.0 in Ry/FeN) at
the theoretical equilibrium lattice constants $a^{th}$, $N(E_F)$ the DOS
at Fermi level in states/eV$\cdot$FeN and $\mu_{Fe}$ the Fe magnetic
moment at the $a^{th}$. $S_{Fe}$ and $S_{N}$ are the radii of Fe and N
atomic spheres corresponding to the theoretical equilibrium volume and
$S_{E}$ the radius of interstitial sphere. }
\begin{center}  \begin{tabular}{cccccccccc}\hline\hline
          &ZnS& &NaCl  &  & & & &CsCl&  \\  \cline{2-2} \cline{4-7} \cline{9-10}
          &NM  & &($00\pi$)-AFM&FM&AFMD&($\pi\pi\pi$)-AFM & &FM&($00\pi$)-AFM  \\ \hline
E$^{th}$ (Ry)&-0.3844& &-0.3901&-0.3892&-0.3856&-0.3799&
&-0.3148&-0.3147\\ $a^{th}$ (\AA) &4.36  &  &4.20 &4.21  &4.19 & 4.20  &
&2.63    &2.63  \\ \hline $S_{Fe}$ (\AA) &1.166& &1.329&1.332&1.326&1.329& &1.425&1.425 \\
$S_{N}$ (\AA)  &0.962& &1.072&1.074&1.070&1.072& &1.129&1.129 \\
$S_{E}$ (\AA)  &1.166& &0.750&0.752&0.749&0.750& &  -   & -     \\
               &0.962& & -   &  -  & -   & -   & & -   &-      \\
$\mu_{Fe}$ ($\mu_{B}$)& - &  &2.67&2.75 &2.67  &2.69  & &   2.86 &2.81 \\
N(E$_F$)    &1.40  & & 1.90  &2.42 &2.39 &1.03  & &1.57  &1.98      \\ \hline
  \end{tabular} \\
{}\end{center}
\end{table}

\clearpage
\baselineskip4ex
\section*{Figure Caption}

\noindent Figure 1: Schematic illustration of FeN with NaCl
(left), ZnS (middle) and CsCl (right) structures.\\

\noindent Figure 2: Calculated total energy vs. volume for FeN with ZnS,
NaCl and CsCl structures. Here ($0,0,\pi$)-AFM and ($\pi,\pi,\pi$)-AFM
denote phases showing antiferromagnetic coupling along [001]- and
[111]-directions, respectively. AFMD is the AFM state which consists of
double layers with FM interlayer coupling along [001]-direction. The
arrows label the volume corresponding to experimental lattice constants of
ZnS- and NaCl-type FeN. \\

\noindent Figure 3: Spin-resolved DOS calculated for FeN with (a)
nonmagnetic ZnS, (b) ($0,0,\pi$)-AFM NaCl and (c) FM CsCl structures at
the theoretical equilibrium volume. Solid lines represent the
majority-spin DOS and dash lines the minority-spin DOS. The Fermi-level is
at zero energy. \\

\noindent Figure 4: Orbital-projected partial DOS calculated for FeN with
(a) NM ZnS, (b) ($0,0,\pi$)-AFM NaCl and (c) FM CsCl structures at
theoretical equilibrium volume. The dash, dot, dash-dotted and solid lines
represent the DOS projected for Fe $s, p, d$ and N $p$ orbitals. The
Fermi-level is at zero energy.\\

\noindent Figure 5: The COHP curves calculated for $s$-$p$,
$p$-$p$ and $d$-$p$ interactions between Fe and N atoms in FeN with (a) NM
ZnS, (b) and (c) ($0,0,\pi$)-AFM NaCl and (d) FM CsCl structures. Here (b)
and (c) show the Fe-N interactions within and between the sublattice in
($0,0,\pi$)-AFM NaCl structure, respectively. The solid, dash-dotted and
dash lines represent $s$-$p$, $p$-$p$ and $d$-$p$ interactions. The integrated
COHP (in eV/cell) for $d$-$p$ interaction are also plotted with thick solid lines.\\

\noindent Figure 6: Calculated Fe magnetic moment in FeN with ZnS, NaCl
and CsCl structures as function of the nearest-neighboring Fe-N distance.
The arrows indicate the distances corresponding to experimental lattice
constants of NaCl- and ZnS-type FeN.

\clearpage
~\\
\vspace{2in}
\begin{center}\begin{figure}[h]
\includegraphics[angle=-90,scale=0.8]{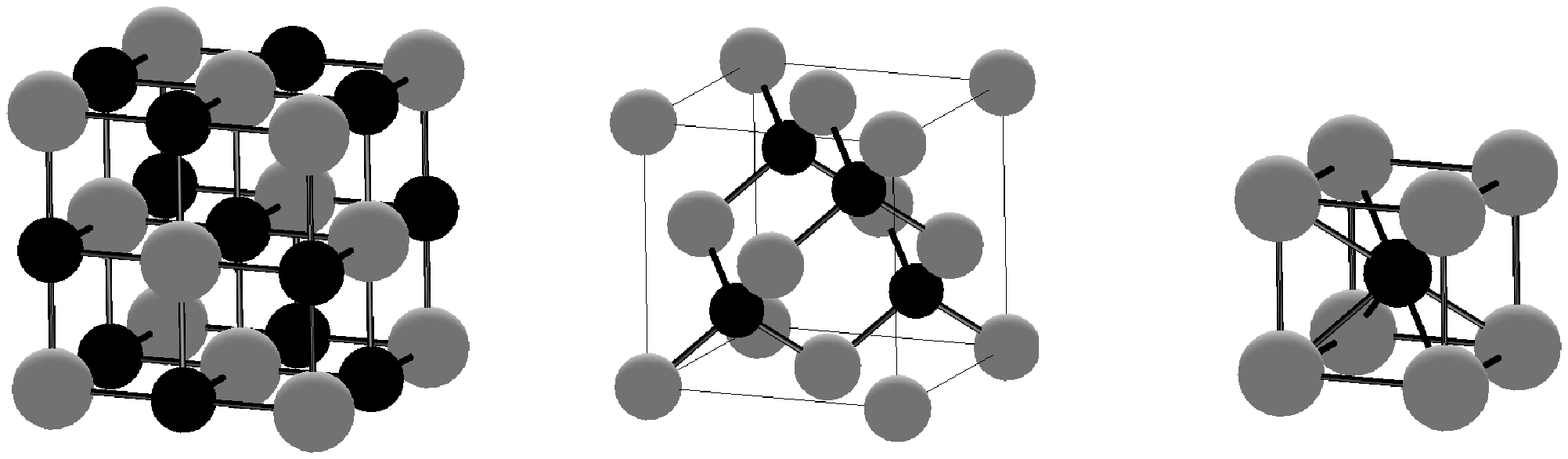}\\
\end{figure}
\vspace{1.5in}\Large Fig. 1 of Y. Kong
\end{center}
%\newpage
\begin{figure}[h]
\includegraphics[bb=80 150 501 709,scale=0.95]{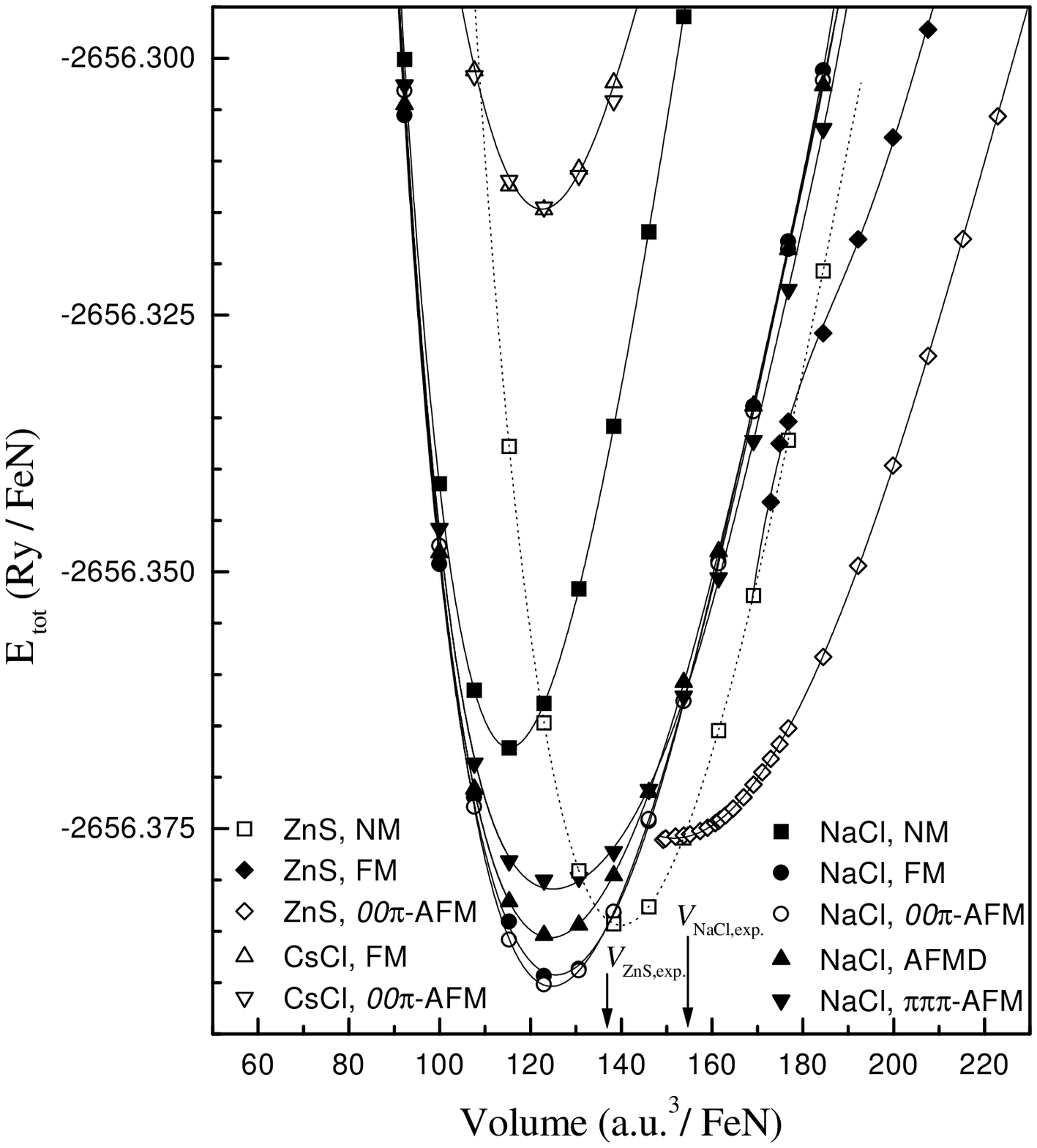}\\
\begin{center}\Large Fig. 2 of Y. Kong\end{center}
\end{figure}
%\clearpage
\begin{figure}[h]
\includegraphics[angle=180,bb=145 108 528 777]{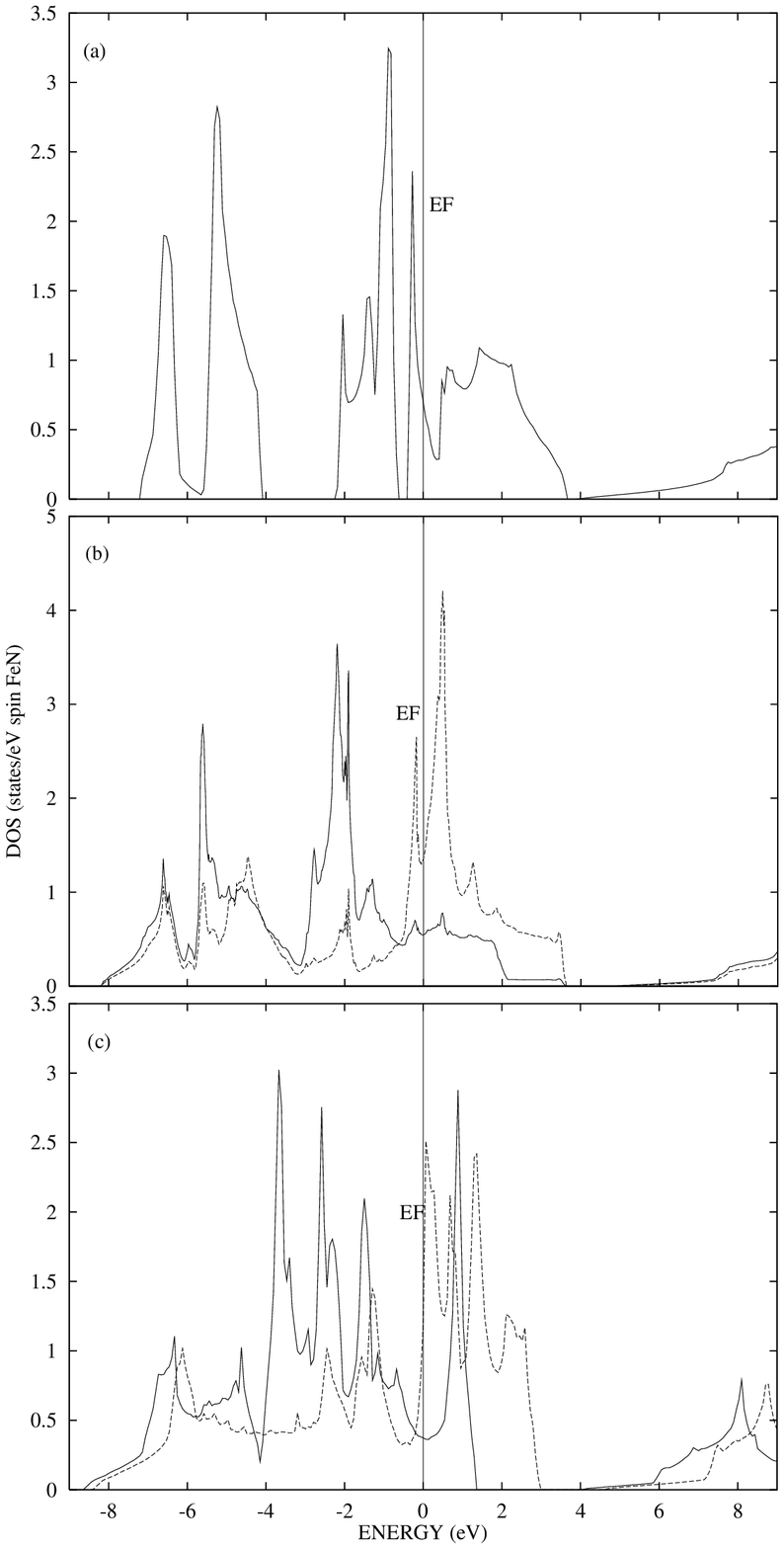}\\
\begin{center}\Large Fig. 3 of Y. Kong\\
\end{center}
\end{figure}
\clearpage
\begin{figure}[h]
\includegraphics[angle=180,bb=145 117 528 777]{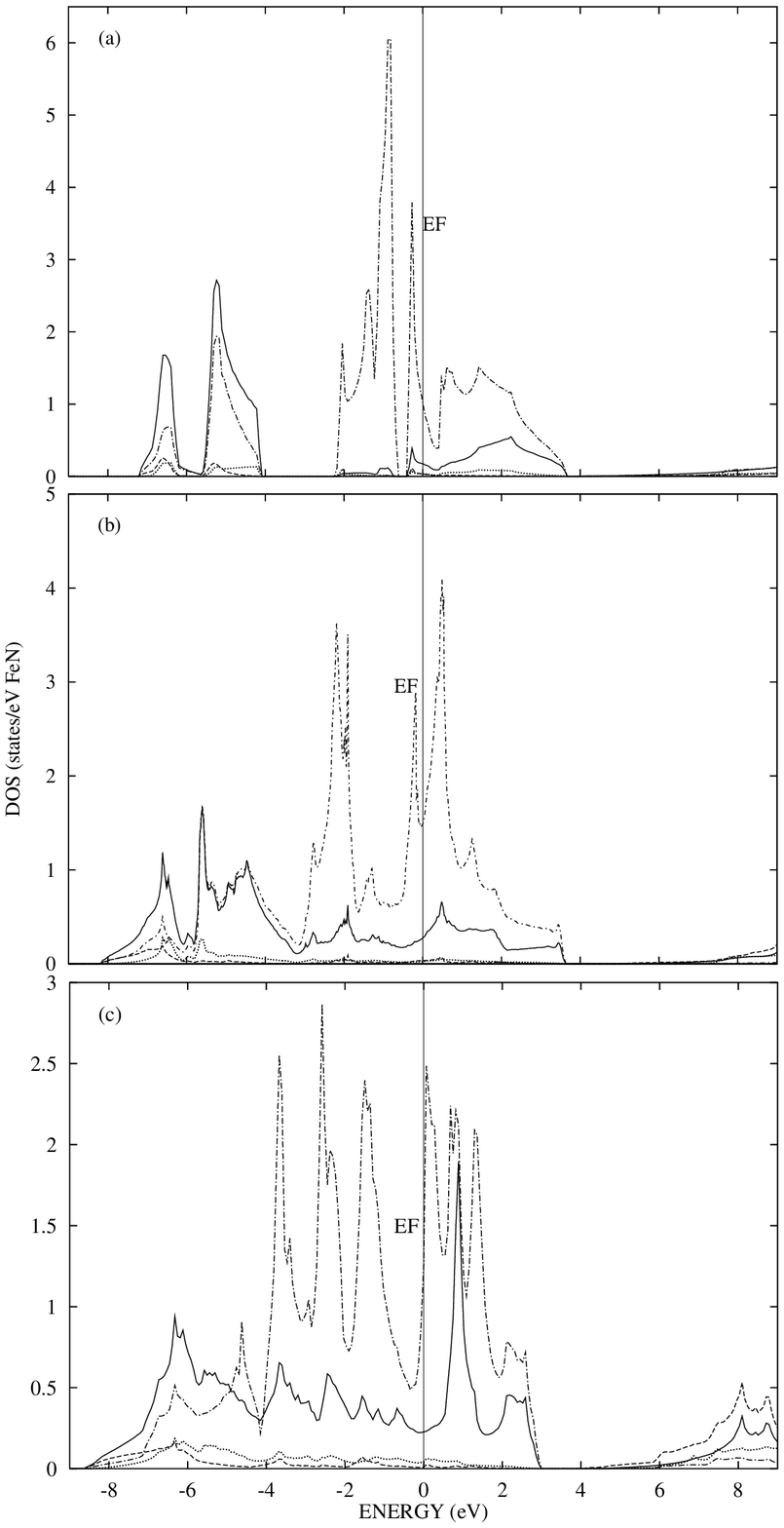}\\
\begin{center}\Large Fig. 4 of Y. Kong\\
\end{center}
\end{figure}
\clearpage
\begin{figure}
\includegraphics[angle=-90,bb=43 98 757 821,scale=0.7]{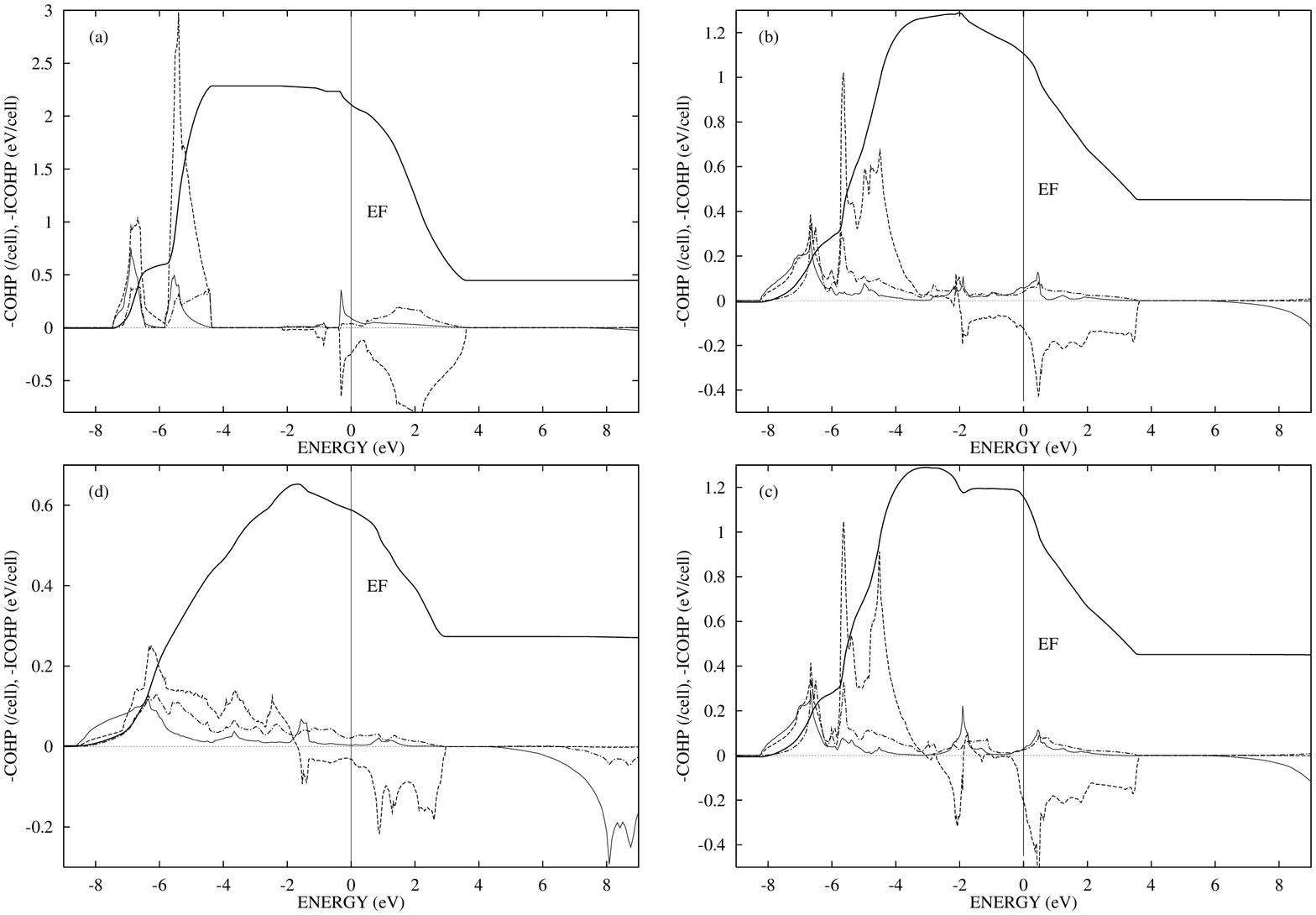}\\
\begin{center}\Large Fig. 5 of Y. Kong\\
\end{center}
\end{figure}
\clearpage
\begin{figure}
\includegraphics[bb=80 150 501 709]{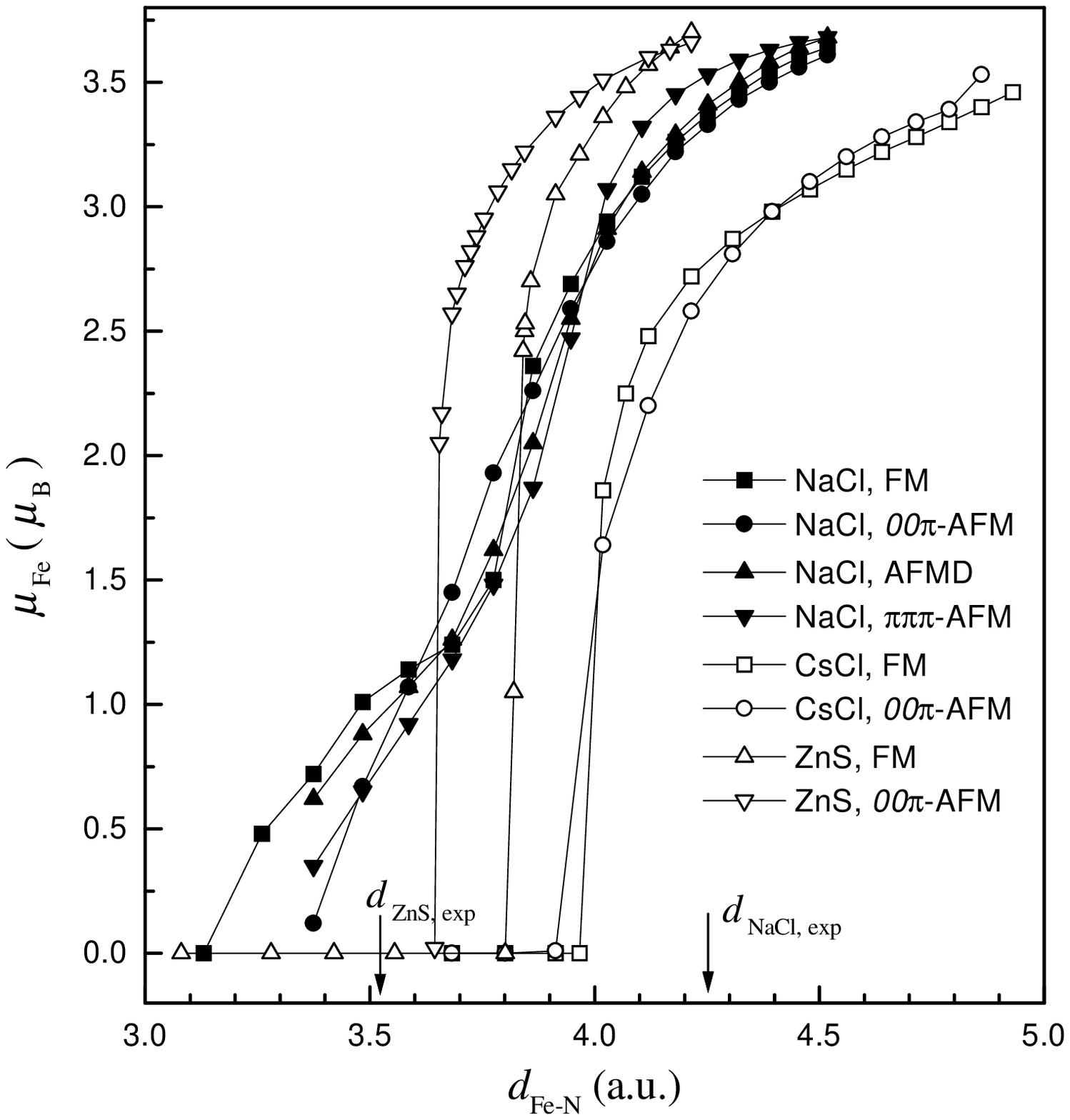}\\
\begin{center}\Large Fig. 6 of Y. Kong\\
\end{center}
\end{figure}

\end{document}